\documentclass[%
 reprint,
 amsmath,amssymb,
 aps,
]{revtex4-2}

\usepackage{graphicx}
\usepackage{dcolumn}
\usepackage{bm}
\usepackage{xcolor} 
\usepackage{multirow}
\usepackage[colorlinks]{hyperref}
\hypersetup{%
	plainpages=true,
	breaklinks=true,
	hypertexnames=false,
	pageanchor=true,
	colorlinks=true,
	linkcolor={blue},
	citecolor={red},
	urlcolor={blue},
	anchorcolor={black}
}
\usepackage{siunitx} 
\usepackage[normalem]{ulem}
\usepackage{hyperref} 
\usepackage{cleveref} 
\usepackage{amsmath}
\usepackage{physics}

\newcommand{\figpanel}[2]{Fig.~\hyperref[#1]{\ref*{#1}(#2)}}
\newcommand{\figpanels}[3]{Fig.~\hyperref[#1]{\ref*{#1}(#2-#3)}}
\newcommand{\figpanelNoPrefix}[2]{\hyperref[#1]{\ref*{#1}(#2)}}
\newcommand{\fullfigpanel}[2]{Figure~\hyperref[#1]{\ref*{#1}(#2)}}
\newcommand{\fullfigpanels}[2]{Figure~\hyperref[#1]{\ref*{#1}(#2)}}

\newcommand{\RNum}[1]{\uppercase\expandafter{\romannumeral #1\relax}}

\usepackage{verbatim}

\begin{document}

\preprint{APS/123-QED}

\title{Deterministic Generation of Frequency-Bin-Encoded Microwave Photons} 

\author{Jiaying Yang$^{1,2}$}
\email{jiyang@chalmers.se}
\author{Maryam Khanahmadi$^{1}$}
\author{Ingrid Strandberg$^{1}$}
\author{Akshay Gaikwad$^{1}$}
\author{Claudia Castillo-Moreno$^{1}$}
\author{Anton Frisk Kockum$^{1}$}
\author{Muhammad Asad Ullah$^{2}$}
\author{Göran Johansson$^{1}$}
\author{Axel Martin Eriksson$^{1}$}
\author{Simone Gasparinetti$^{1}$}
\email{simoneg@chalmers.se}
\homepage{https://202q-lab.se}

\affiliation{$^1$Department of Microtechnology and Nanoscience, Chalmers University of Technology, SE-412 96, G\"{o}teborg, Sweden
\\$^2$ Ericsson Research, Ericsson AB, SE-164 83, Stockholm, Sweden
}

\date{\today}

\begin{abstract}
A distributed quantum computing network requires a quantum communication channel between spatially separated processing units. In superconducting circuits, such a channel can be implemented based on propagating microwave photons to encode and transfer quantum information between an emitter and a receiver. However, traveling microwave photons can be lost during the transmission, leading to the failure of information transfer. Heralding protocols can be used to detect such photon losses. In this Letter, we propose such a protocol and experimentally demonstrate a frequency-bin encoding method of microwave photonic modes using superconducting circuits. We deterministically encode the quantum information from a superconducting qubit by simultaneously emitting its information into two photonic modes at different frequencies, with a process fidelity of \SI{94.9}{\percent}. 
The frequency-bin-encoded photonic modes can be used, at the receiver processor, to detect the occurrence of photon loss. Our Letter thus provides a reliable method to implement high-fidelity quantum state transfer in a distributed quantum computing network, incorporating error detection to enhance performance and accuracy. 
\end{abstract}

\maketitle

\textit{Introduction---} Distributed quantum computing architectures, which allow multiple quantum processors to collaborate on larger tasks, offer a promising approach to improving the scalability of quantum computation~\cite{Cirac1997quantum, kimble2008quantum}. Implementing these architectures requires establishing quantum channels between remote quantum processors. 
For trapped-ion qubits~\cite{PhysRevLett.130.050803}, neutral atoms~\cite{doi:10.1126/science.1221856, ritter2012elementary}, semiconductor quantum dots~\cite{delteil2016generation, Phase2017Stockill}, and color-center qubits~\cite{bernien2013heralded, humphreys2018deterministic}, quantum channels can be implemented using optical fibers to transmit photons over long distances. 
For superconducting circuits, quantum channels at microwave frequencies have been created using standing modes~\cite{leung2019deterministic, zhong2021deterministic} or traveling microwave photons~\cite{Campagne-Ibarcq2018deterministic, kurpiers2018deterministic, axline2018demand, Magnard2020microwave, qiu2023deterministic, PhysRevLett.132.047001}, with the latter enabling a photon carrying quantum information to be emitted by the sender processor, travel through the quantum channel, and be reabsorbed by the receiver processor. This process enables deterministic quantum state transfer and remote entanglement between the qubits on both processors.

However, the emitted microwave photons are exposed to noise during the transmission in the quantum channel, posing a risk of photon loss or other signal corruption when transmitting quantum information between remotely distributed quantum processors. Heralding protocols for error detection can eliminate this issue by introducing an additional degree of freedom into the emission, such as time-bin~\cite{ lo2020quantum, saha2024high}, frequency-bin~\cite{Lukens:17, lu2020fully, lu2023frequency}, polarization~\cite{Peters2005Remote}, path~\cite{Kok2002creation, Matthews2011heralding}, or angular-momentum encoding~\cite{Yao:11}, which are commonly used in the optical regime.  In the microwave regime, time-bin-encoded photons have been experimentally demonstrated using superconducting qubits~\cite{ilves2020demand, kurpiers2019quantum, li2023frequency}. By emitting photons at different frequencies simultaneously, frequency-bin encoding offers a higher data rate, effectively doubling the transmission speed compared to time-bin encoding using the same device. However, to the best of our knowledge, no implementation of the frequency-bin-encoded photon exists in the microwave regime. Furthermore, if such an implementation were to be developed, it would typically require additional hardware resources, such as extra qubits functioning as quantum state emitters, to emit both photonic modes simultaneously.

In this Letter, we propose and experimentally demonstrate frequency-bin-encoded photon generation, utilizing superconducting qubits in a waveguide quantum electrodynamics setup. The transmitted quantum information is encoded into a pair of simultaneously emitted propagating temporal modes at different frequencies, rather than between the vacuum state and the Fock state \( \left| 1 \right\rangle \). The fact that the vacuum is not a logical state allows the frequency-bin-encoded photon to serve as a heralding protocol for error detection, addressing photon loss at the receiver processor. Our implementation is based on a qubit-coupler-qubit configuration~\cite{besse2020realizing, kannan2023demand, yang2023deterministic}. However, instead of using the coupler to mediate qubit interactions, we tune the coupler into resonance with the emitter qubit, forming a hybridized emitter. By applying simultaneous excitation-preserving and nonpreserving control pulses, we transfer the quantum state to the hybridized emitter and generate frequency-bin-encoded photons. A straightforward extension of the qubit-coupler-qubit configuration would require an extra emitter qubit and two extra couplers to generate photonic modes at two frequencies [see Supplemental Material~\cite{Sup} Fig.~S1]. In contrast, our approach leverages hybridized modes and thus omits those additional components, making frequency-bin encoding more hardware‐efficient.  
We perform joint quantum state tomography on the frequency-bin modes, demonstrating encoding of qubit information into the frequency bins with an average state fidelity of \SI{97.5}{\percent} and a process fidelity of \SI{94.9}{\percent} in this first, proof-of-principle realization. 
Our device can serve as a building block in distributed quantum computing systems, functioning as either an emitter or a receiver.

\begin{figure}
\includegraphics[width=0.99\linewidth]{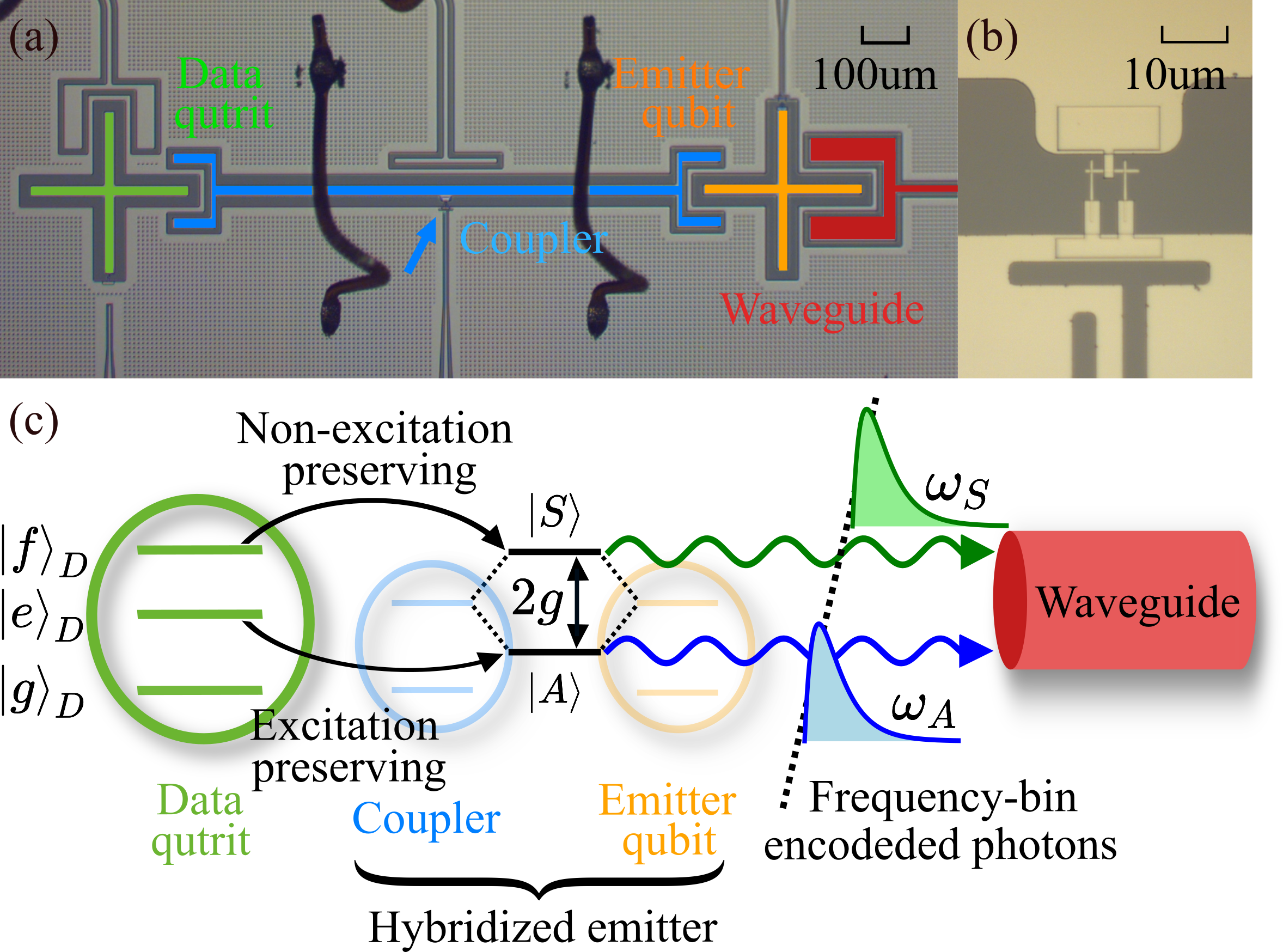}
\caption{Experimental generation of frequency-bin-encoded photons via a superconducting circuit. (a) False-color micrograph showing a data qutrit (green), emitter qubit (orange), flux-tunable coupler (blue) with airbridges (black) for robust grounding, and a waveguide (red). (b) Enlarged view of the coupler's SQUID, whose position is indicated by a blue arrow in (a). (c) Schematic representation of the device. The two frequency-encoded photonic modes are emitted at different frequencies $\omega_A$ (blue) and $\omega_S$ (green) simultaneously.
}
\label{sample}
\end{figure}

\textit{Experimental setup--- } Our device is a superconducting quantum circuit comprising two Xmon-type~\cite{barends2013coherent} transmon qubits~\cite{koch2007charge}—a data qutrit (D) and an emitter qubit (E)—capacitively coupled via a parametric coupler~\cite{mckay2016universal} [\figpanel{sample}{a}]. Qutrit D has a transition frequency of \(\omega_{D}^{ge} / 2\pi = \SI{5.05}{GHz}\) between its ground state \(\left| g \right\rangle\) and first excited state \(\left| e \right\rangle\). We consider the three-level subspace of Qutrit D, including also the second excited state \(\left| f \right\rangle\), with an anharmonicity of $\alpha/2\pi = -\SI{215}{MHz}$. Both the coupler and qubit E are frequency-tunable by on-chip flux lines, enabled by superconducting quantum interference devices (SQUIDs). Their maximum frequencies are $\omega_{C,0}/2\pi=\SI{8.46}{GHz}$ and $\omega_{E,0}/2\pi=\SI{6.17}{GHz}$, respectively, at zero magnetic flux. Qubit E is capacitively coupled to a coplanar waveguide with a decay rate of $\Gamma_{E}/2\pi = \SI{8}{MHz}$. Qutrit D and the coupler are also capacitively coupled to one $\lambda/4$ coplanar waveguide resonator each for characterization and readout. A summary of measured circuit parameters is available in Supplemental Material~\cite{Sup} Sec.~\RNum{2}.

\textit{Hybridized emitter coupled to the waveguide---} We generate frequency-bin-encoded photonic modes by simultaneously emitting two modes at distinct frequencies using the arrangement shown in \figpanel{sample}{c}. 
With a parametric coupler connected to Qubit E, and Qubit E being the only qubit physically coupled to the waveguide,  
we tune the coupler and Qubit E into resonance. 
The resonance between the coupler and Qubit E results in a hybridized emitter with two modes~\cite{PhysRevA.84.061805, Aamir2022symmetryselective}, a symmetric state $\left | S \right \rangle = (\left | g \right \rangle_C\left | e \right \rangle_E +\left | e \right \rangle_C\left | g \right \rangle_E )/\sqrt 2$ at $\omega_S/2\pi = \SI{5.79}{GHz}$ and an antisymmetric state $\left | A \right  \rangle =(\left | g \right \rangle_C\left | e \right \rangle_E -\left | e \right \rangle_C\left | g \right \rangle_E )/\sqrt 2$ at $\omega_A/2\pi = \SI{5.70}{GHz}$, where $\ket{g,e}_{C(E)}$ define the bare states of the coupler (Qubit E) in a single-excitation manifold. 
The splitting between the two hybridized modes is $(\omega_S - \omega_A)/2\pi = 2g/2\pi = \SI{92}{MHz}$, with $g$ the coupling rate between the coupler and Qubit E. These hybridized modes are coupled with the same strength to the waveguide, enabling simultaneous emission without requiring an additional degree of freedom, which leads to hardware-efficient operation. 
We apply an excitation-preserving parametric drive~\cite{mckay2016universal, besse2020realizing, yang2023deterministic} through the coupler to transfer the first excited state of Qutrit D to the antisymmetric mode, and a nonexcitation-preserving second-order-transition drive~\cite{Pechal2014microwave, kurpiers2018deterministic, kurpiers2019quantum, ilves2020demand} through the charge line on Qutrit D, to transfer its second excited state to the symmetric mode. Thus, the hybridized emitter can be utilized as a source to emit two distinct propagating modes at their respective frequencies.

\begin{figure}\centering
\includegraphics[width=0.96\linewidth]{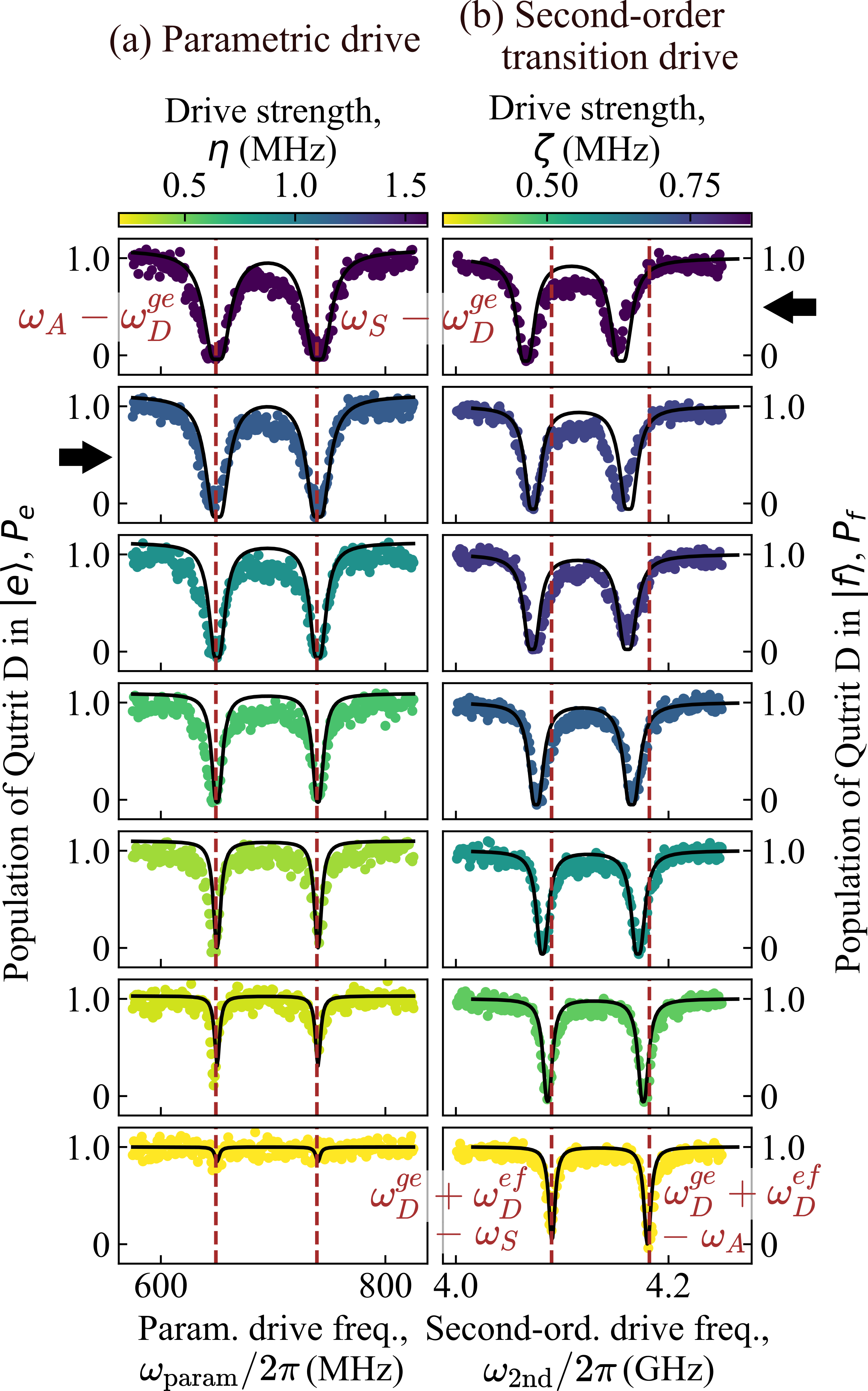}
\caption{Characterization of the hybridized modes. (a) The population of Qutrit D in state \(\left | e \right \rangle_D\) measured while sweeping the frequency and amplitude of the parametric drive; (b) the population of Qutrit D in state \(\left | f \right \rangle_D\) measured while sweeping the frequency and amplitude of the second-order-transition drive. In both panels, the filled circles indicate the measured data and the solid black curves represent the corresponding numerical simulation (with the drive strengths used in subsequent experiments highlighted by black arrows).}
\label{hybrid-modes}
\end{figure}

To characterize the population transfer to the hybridized emitter, we perform two measurements with varying drive strengths ($\eta, \zeta$) and frequencies ($\omega_{param}, \omega_{2nd}$) of the parametric and second-order-transition drives. First, with Qutrit D initialized in state \(\left | e \right \rangle_D\), we measure its population while sweeping the parametric-drive frequency $\omega_{\rm param}$. When the frequency matches the resonance conditions $\omega_A-\omega_D^{ge}$ or $\omega_S-\omega_D^{ge}$, the population transfers to the hybridized modes, reducing the population in $\left | e \right \rangle_D$ [\figpanel{hybrid-modes}{a}]. The population dips become deeper and broader as the parametric-drive amplitude increases. In the second measurement, with Qutrit D initialized in \(\left | f \right \rangle_D\), we sweep the second-order drive frequency \(\omega_{\rm 2nd}\). Resonant frequencies \(\omega_{\rm 2nd} = \omega_D^{ge} + \omega_D^{ef} - \omega_S\) and \(\omega_{\rm 2nd} = \omega_D^{ge} + \omega_D^{ef} - \omega_A\) result in dips in the population of \(\left | f \right \rangle_D\) [\figpanel{hybrid-modes}{b}], indicating the transfer of excitation to the waveguide via the hybridized modes. The frequency shift of the hybridized modes, induced by the parametric drive and the nonlinear relation between the coupler frequency and the dc flux~\cite{mckay2016universal, PhysRevA.96.062323}, is small, as evidenced by the near-constant dip frequencies despite increasing pulse amplitude [\figpanel{hybrid-modes}{b}]. In contrast, the ac Stark shift from the second-order drive~\cite{kurpiers2018deterministic} is significantly larger, as seen by the spectroscopy dips shifting to the left with increasing amplitude [\figpanel{hybrid-modes}{c}], which needs to be considered in the following experiments. 

\begin{figure}\centering
\includegraphics[width=0.8\linewidth]{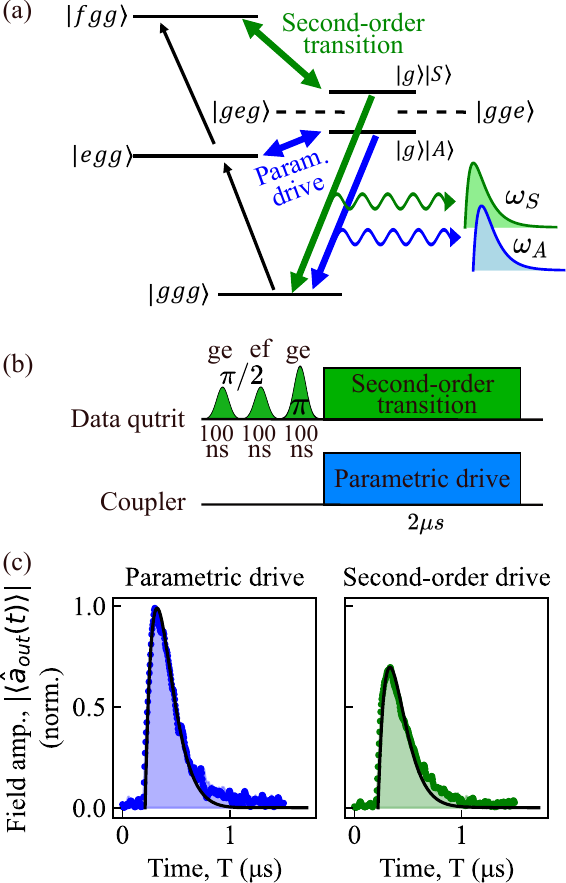}
\caption{Emitted frequency-bin-encoded photonic modes. (a) Energy-level diagram of the emission process; \( |ijk\rangle \) (\(i,j,k=g,e,f\)) denote the states of Qutrit D, the coupler, and Qubit E, respectively. (b) Pulse sequence for generating a photon with nonzero amplitude. (c) Amplitude of the frequency-bin-encoded photonic modes. The filled circles represent the measured photon magnitude when the photonic modes are emitted simultaneously using the parametric and second-order-transition drives. The shaded area represents the measured photon magnitude when each drive pulse is applied individually, and each photon is measured separately. The solid black curves represent the theoretical prediction from numerical simulations.
}
\label{emit-photon}
\end{figure}

\textit{Generation of frequency-bin-encoded photons.---} The frequency-bin-encoded photon is generated by simultaneously driving the transitions $\left | e \right \rangle_D \to \left | A \right \rangle$ and $\left | f \right \rangle_D \to \left | S \right \rangle$ using the parametric and second-order drives, respectively [\figpanel{emit-photon}{a}]. We start by initializing Qutrit D in an arbitrary superposition state of the ground and first excited states, $\alpha \left | g \right \rangle_D + \beta \left | e \right \rangle_D$. Next, $\pi_{ef}$ and $\pi_{ge}$ pulses transform this state into the same superposition between the first and second excited states, $\alpha \left | e \right \rangle_D + \beta \left | f \right \rangle_D$. Subsequently, we simultaneously apply the parametric and second-order-transition drives. Because of the Stark-shift frequency $\omega_{\text{ac}}$ from the second-order drive, we fix the frequency of the parametric drive and the second-order transition as $\omega_{\mathrm{param}} + \omega_{\text{ac}}$ and $\omega_{\rm 2nd} + 2\omega_{\text{ac}}$, 
which transfer the population to the $\left | A \right \rangle$ and $\left | S \right \rangle$ states of the hybridized emitter, respectively, yielding the superposition \(\alpha \left | A \right \rangle + \beta \left | S \right \rangle\).
%
Because the hybridized emitter has a large decay rate into the waveguide ($\Gamma \gg \eta, \zeta$), it immediately decays to the ground state, emitting the entangled propagating modes \(\alpha \left | 1 \right \rangle_{\omega_A} \left | 0 \right \rangle_{\omega_S} + \beta \left | 0 \right \rangle_{\omega_A} \left | 1 \right \rangle_{\omega_S}\) into the waveguide, where \(i\) and \(j\) in \(\left | i \right \rangle_{\omega_A} \left | j \right \rangle_{\omega_S}\) denote the photon numbers at frequencies \(\omega_A\) and \(\omega_S\), respectively. See End Matter and Supplemental Material~\cite{Sup} for more details about the system Hamiltonian.

It is worth noting that for each frequency mode obtained from the quantum state \(\ket{\psi}=\alpha \left | 1 \right \rangle_{\omega_A} \left | 0 \right \rangle_{\omega_S} + \beta \left | 0 \right \rangle_{\omega_A} \left | 1 \right \rangle_{\omega_S},\,\, \forall \, \alpha,\beta\), the reduced density matrix has no coherences in the Fock basis, \textit{i.e.}, $\bra{0}\text{Tr}_{A(S)}[\ket{\psi}\bra{\psi}]\ket{1}=0$ (see Supplemental Material~\cite{Sup} Fig.~S4 for examples of reduced density matrices).
In other words, the radiation emitted into each of the modes consists of a statistical mixture of vacuum and single photons; as such, it has a random phase and an amplitude averaging to zero. To characterize the emitted radiation with amplitude measurements, we thus introduce a variation of the pulse sequence~[\figpanel{emit-photon}{b}]. 
We introduce the vacuum energy level $ \ket{0}_{\omega_A} \ket{0}_{\omega_S}$ to the state by applying two $\pi/2$ pulses and a $\pi$ pulse, followed by the two drives to obtain the propagating state $\frac{1}{2} \ket{0}_{\omega_A} \ket{0}_{\omega_S}  +\frac{1}{\sqrt{2}} \ket{1}_{\omega_A} \ket{0}_{\omega_S} +\frac{1}{2} \ket{0}_{\omega_A} \ket{1}_{\omega_S}$. The emitted photon amplitude of the two frequency modes has a fast rise followed by an exponential decay [\figpanel{emit-photon}{c}], where the speed of the photon emission, or the effective decay rate, is controlled by adjusting the drive strengths $\eta$ and $\zeta$. Parametric emission is theoretically expected to have an amplitude $\sqrt 2$ times that of second-order-transition emission, due to the state's amplitude coefficients. This amplitude ratio between the two modes, shown in \figpanel{emit-photon}{c}, is derived from the moment normalization as explained in Supplemental Material~\cite{Sup} Sec.~\RNum{4}-C.

\begin{figure*}\centering
\includegraphics[width=0.93\linewidth]{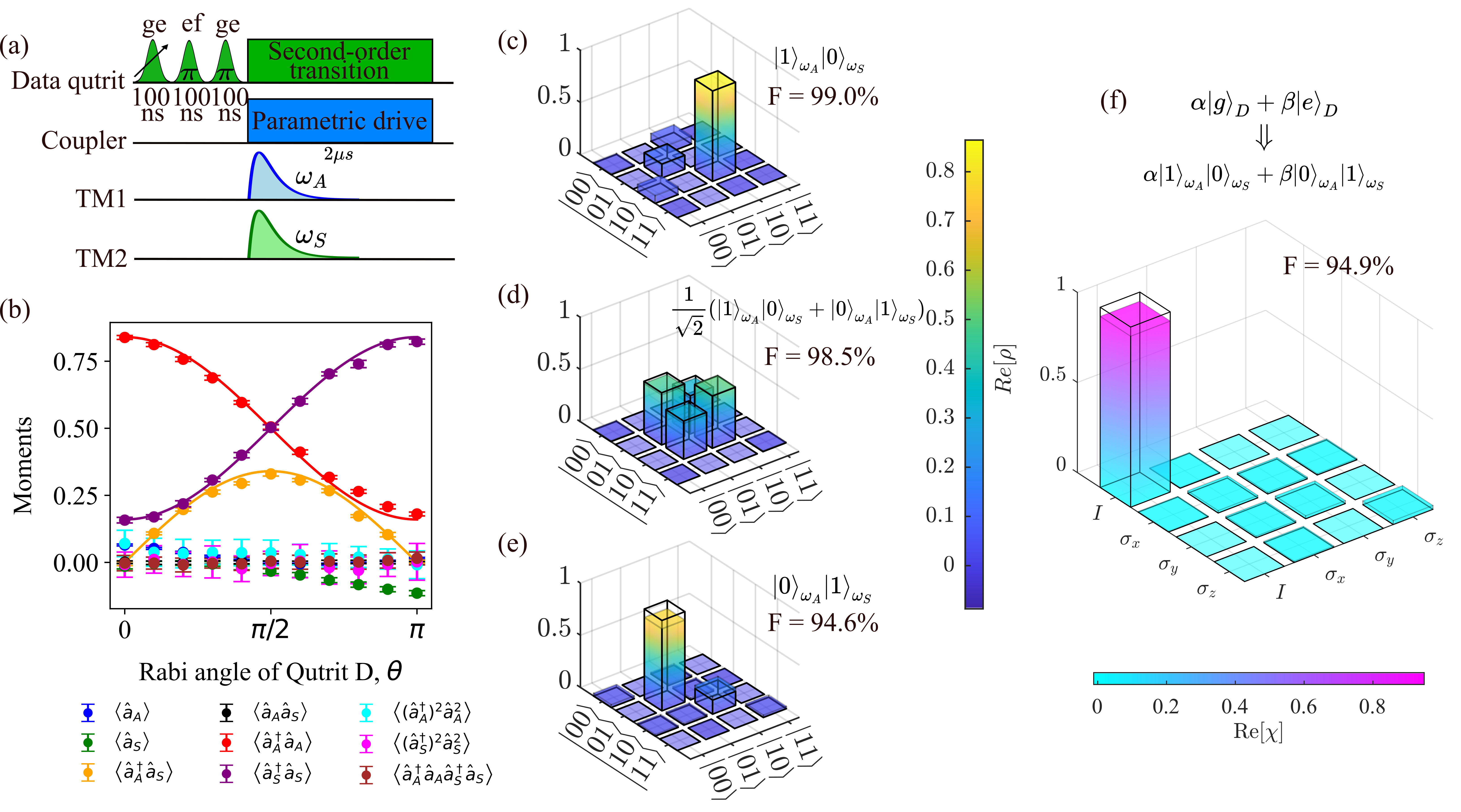} 
\caption{Tomography measurement of the frequency-bin-encoded photon output field. 
(a) Pulse sequence for the tomography measurement. TM1 and TM2 are the two temporal filters for temporal mode matching.
(b) 
Selected moments of the emitted photon field. Filled circles are the measured data, which account for the thermal population $P_{\mathrm{thermal}}$ in the normalization. The solid curves show the expected values of the corresponding-colored moments, also incorporating $P_{\mathrm{thermal}}$. Moments without a solid curve in the same colors have zero expected value for any Rabi angle $\theta$. 
(c-e) Reconstructed (colored bars) and expected (black wireframes, including thermal population) density matrices of the frequency-bin-encoded propagating modes for Qutrit D initialized in states (c) $\left | g \right \rangle_D$, (d) $\frac{1}{\sqrt{2}} (\left | g \right \rangle_D + \left | e \right \rangle_D)$, and (e) $\left | e \right \rangle_D$. The fidelity \(F\) is indicated next to each density matrix. The imaginary elements of the matrices, all below 0.1, are omitted. 
(f) The process matrix $\chi$ for state transfer between Qutrit D and the propagating mode, obtained from quantum process tomography measurements. The colored bars represent the measured results, while the black wireframes depict the ideal process. The imaginary elements of $\chi$, all below 0.03, are omitted.  
}
\label{tomography}
\end{figure*}

\textit{Characterization of frequency-bin-encoded photons---} We characterize the photonic modes through temporal mode matching~\cite{eichler2011experimental} and joint quantum state tomography. Using temporal mode matching, we select the desired propagating modes $\hat{a}_A$ and $\hat{a}_S$ from the continuum of modes in the waveguide, following the method described in Ref.~\cite{eichler2012characterizing}. The moments of the two photonic modes are measured when Qutrit D is initialized in the state \(\alpha \left | g \right \rangle_D + \beta \left | e \right \rangle_D\) for different values $(\alpha,\beta)$ achieved by sweeping the Rabi angle $\theta$, where $\alpha = \cos(\theta/2)$ and $\beta = \sin(\theta/2)$ [see \figpanel{tomography}{a} for the pulse sequence and \figpanel{tomography}{b} for the moments; detailed explanations are provided in the End Matter]. 

By initializing Qutrit D in the states $\left |g \right \rangle_D$, $\frac{1}{\sqrt{2}} (\left |g \right \rangle_D + \left |e \right \rangle_D)$, and $\left |e \right \rangle_D$, 
and following the same pulse sequence to perform joint quantum state tomography, we reconstruct the density matrices of the frequency-bin-encoded photonic modes for the three cases [\figpanels{tomography}{c}{e}]. In the reconstruction, we use least-squares optimization~\cite{PhysRevResearch.2.042002, PhysRevApplied.18.044041} to find the most likely density matrix of the frequency-bin-encoded photonic modes corresponding to the measured moments. 
The fidelity of the density matrix is defined as $F(\rho_{\rm in}, \rho) = \left(\mathrm{Tr}\sqrt{\sqrt{\rho_{\rm in}}\rho\sqrt{\rho_{\rm in}}}\right)^2$~\cite{jozsa1994fidelity}, with $\rho_{\rm in}$ being the expected density matrix and $\rho$ being the measured one. To account for Qutrit D’s thermal population at all preparation Rabi angles, we incorporate a nonzero excited-state population  $P_{\mathrm{thermal}}$ into $\rho_{\rm in}$. Specifically, rather than assuming a pure excited state at zero Rabi angle, we include an initial thermal population, after which we apply the ideal unitary rotations for other Rabi angles to obtain $\rho_{\rm in}$. 
The obtained fidelities of the density matrices are \SI{99.0}{\percent}, \SI{98.5}{\percent}, and \SI{94.6}{\percent}, 
for the three photonic states $\left |1\right \rangle_{\omega_A} \left |0\right \rangle_{\omega_S}$, $\frac{1}{\sqrt{2}} (\left |1\right \rangle_{\omega_A} \left |0\right \rangle_{\omega_S}+\left |0\right \rangle_{\omega_A}\left |1\right \rangle_{\omega_S})$, and $\left |0\right \rangle_{\omega_A}\left |1\right \rangle_{\omega_A}$, respectively. The simulated density matrices yield fidelities of \SI{99.2}{\percent}, \SI{99.0}{\percent}, and \SI{98.7}{\percent}, accounting for Qutrit D and coupler relaxation and thermal excitation. See Supplemental Material~\cite{Sup} Sec.~\RNum{4} for fidelities computed with respect to the ideal case without thermal excitations. 

The quantum process tomography is done by initializing Qutrit D in one more state, $\frac{1}{\sqrt{2}}(|g\rangle_D-i|e\rangle_D)$, in addition to the three states above, which correspond to the cardinal states on the Bloch sphere. We then utilize the reconstructed density matrices of the frequency-bin-encoded photons for these four states, along with least-squares optimization, to determine the process matrix $\chi$~\cite{gaikwad-sr-2022}. The process fidelity is found to be \SI{94.9}{\percent} 
[\figpanel{tomography}{e}] (see Supplemental Material~\cite{Sup} Sec.~\RNum{4} for details on both state and process tomography).

\textit{Discussion---} In this work, we experimentally demonstrated the generation of frequency-bin-encoded photonic modes in microwave regime. 
By leveraging the hybridized emitter formed by the coupler and Qubit E, and applying both excitation-preserving and non-preserving drives simultaneously, we achieved two-frequency-mode emission from a single qubit physically coupled to a waveguide. 
Additionally, this setup enables photon emission both with and without frequency-bin encoding from the same device, allowing us to encode quantum information into photonic modes while maintaining the flexibility to integrate or bypass the error-detection protocol using the same system. Decoherence theoretically limits the average state fidelity to \SI{99.0}{\percent}, but our measurements show \SI{97.5}{\percent}. This gap is primarily due to coherent control pulse errors between Qutrit D and the hybridized emitter, which can be reduced with optimal control techniques~\cite{motzoi2009simple,werninghaus2021leakage}. The remaining infidelity arises from incoherent errors, including qubit and coupler decoherence. Qutrit D suffers from thermal excitation, which can be mitigated through active or passive qubit reset~\cite{magnard2018fast, aamir2025thermally}. This frequency-bin-encoded photon generation method is also applicable to other types of solid-state quantum hardware, such as quantum dots~\cite{stockklauser2017}. 

To demonstrate a distributed quantum computing architecture with frequency-bin-encoded photons, we can reshape the envelopes of the photonic modes to be time-symmetric as in Ref.~\cite{Pechal2014microwave, yang2023deterministic}, and use the same structure for photon reabsorption, in a time-reversed way. 
In quantum state transfer, frequency-bin-encoded photonic modes serve as an error-detection protocol to identify photon loss and trigger the information retransmission (see Supplemental Material~\cite{Sup} Sec.~\RNum{5} for the state transfer process and End Matter for error-detection details). Remote entanglement can be achieved by adding an extra qubit to the emitter node (see Supplemental Material~\cite{Sup} Sec.~\RNum{5}), which also facilitates error detection. Our work thus forms a foundational building block for distributed quantum computing architectures and quantum networks.


\textit{Acknowledgments---}
The authors thank Mohammed Ali Aamir for useful discussions, and Lars Jönsson for his help in making the sample holders. The device in this work was fabricated in Myfab, Chalmers, a micro- and nanofabrication laboratory. The traveling-wave parametric amplifier (TWPA) used in this experiment was provided by IARPA and Lincoln Laboratory. This work was supported by Ericsson Research and the Knut and Alice Wallenberg Foundation through the Wallenberg Centre for Quantum Technology (WACQT). S.G. acknowledges financial support from the European Research Council via Grant No. 101041744 ESQuAT. A.F.K. is also supported by the Swedish Research Council (Grant No. 2019-03696), the Swedish Foundation for Strategic Research (Grant No. FFL21-0279 and No. FUS21-0063), and the Horizon Europe program HORIZON-CL4-2022-QUANTUM-01-SGA via the project No. 101113946 OpenSuperQPlus100.

S.G. planned the project. J.Y. performed the experiments and analyzed the data with input from A.M.E., I.S., S.G., and M.A.U.. M.K. and G.J. built the theoretical model. J.Y. designed the device and C.C.M. fabricated the device.  A.G. and A.F.K. operated the quantum state tomography and quantum process tomography. J.Y. and M.K. wrote the manuscript with feedback from all authors. S.G. and A.M.E. supervised this work. 

\bibliography{main}

\section*{End Matter}
\textit{Appendix A: Effective Hamiltonian ---}
Here, we detail the theoretical framework underlying our photon-emission process via the hybridized emitter. In a rotating frame defined by the Hamiltonian $H_0 = \omega_D^{ge} \hat d^\dagger \hat d + \omega_A \hat a_A^\dagger \hat a_A + \omega_S \hat a_S^\dagger \hat a_S$, the driven system is described by the effective Hamiltonian
\begin{align}\label{hamiltonian}
    H_{\mathrm{eff}} = (\eta\hat d^\dagger \hat a_A+ \eta^*\hat d\hat a_A^\dagger) +(\zeta\hat d^{\dagger 2}\hat a_S+\zeta^*\hat d^2\hat a_S^\dagger) +\frac{\alpha}{2} \hat d^{\dagger 2}\hat d^{ 2}\, ,
\end{align}
 where the ladder operators $\{(\hat d,\hat d^\dagger),(\hat a_A, \hat a_A^\dagger),(\hat a_S,\hat a_S^\dagger)\}$ describe the mode of Qutrit D, the symmetric mode, and the antisymmetric mode, respectively (see Supplemental Material~\cite{Sup} Sec.~\RNum{3} for the full Hamiltonian). We do not consider the higher excited states of the coupler and Qubit E since they are barely occupied due to the anharmonicity. The drive strengths of the parametric drive ($\eta$) and the second-order transition ($\zeta$) should satisfy the relation $\eta = \sqrt{2} \zeta$, to ensure approximately equal emission rates from Qutrit D to the two propagating modes, as the two drives correspond to one- and two-photon transitions, respectively. 
 In our case, the strengths are chosen to be $|\eta |/2\pi= \SI{1.12}{MHz}$ and $|\zeta|/2\pi= \SI{.79}{MHz}$ for the photon emission and tomography experiments. 
 
\textit{Appendix B: The extended description of moments ---} As mentioned in the main text, the first-order moments representing the mean amplitude of the field,  $\langle\hat{a}_A\rangle$ and $\langle\hat{a}_S\rangle$, are always zero because the reduced density matrix of each mode has no coherences. The mean photon numbers of the field are represented by $\langle\hat a_A^\dagger \hat a_A\rangle$ and $\langle\hat a_S^\dagger \hat a_S\rangle$, reaching the maximum values when Qutrit D is initialized in $\left |g\right \rangle_D$ and $\left |e\right \rangle_D$, respectively. \(\langle\hat{a}_S^\dagger \hat{a}_S\rangle\) is expected to be zero when the Rabi angle \(\theta\) is \(0\); however, it is finite due to the thermal excitation of Qutrit D. Similarly, \(\langle\hat{a}_A^\dagger \hat{a}_A\rangle\) should be zero when \(\theta\) is \(\pi\), but it remains nonzero because of the themal excitation of Qutrit D. In the first case, all population of Qutrit D is transferred into the photon state $\left | 1 \right \rangle_{\omega_A} \left | 0 \right \rangle_{\omega_S}$, while in the latter case, all population of Qutrit D is transferred into the state $\left | 0 \right \rangle_{\omega_A} \left | 1 \right \rangle_{\omega_S}$. The cross-correlated second-order moment of the two modes, $\langle\hat a_A^\dagger \hat a_S\rangle$, signals entanglement between the two photonic modes, which reaches the maximum at $\theta = \pi/2$ when we get an equal superposition state between the two modes, $\frac{1}{\sqrt{2}} (\left | 1 \right \rangle_{\omega_A} \left | 0 \right \rangle_{\omega_S}+ \left | 0 \right \rangle_{\omega_A} \left | 1 \right \rangle_{\omega_S})$. The other cross-correlated moment $\left<\hat a_A \hat a_S\right>$ is zero for all Rabi angles, because the two photons are never emitted at the same time. The fourth-order moments $\langle(\hat a_A^{\dagger})^2\hat a_A^2\rangle$ and $\langle(\hat a_S^{\dagger})^2\hat a_S^2\rangle$ are expected to be zero for all $\theta$ since maximally a single photon is emitted in each mode. The cross-correlated fourth-order moment $\langle\hat a_A^{\dagger}\hat a_A\hat a_S^{\dagger} \hat a_S\rangle$ is also zero, as the information is encoded into the superposition state $\alpha\left | 1 \right \rangle_{\omega_A} \left | 0 \right \rangle_{\omega_S}+ \beta\left | 0 \right \rangle_{\omega_A} \left | 1 \right \rangle_{\omega_S}$ for any Rabi angle $\theta$, where the absence of simultaneous photon occupation in both modes ensures that no joint detection is possible. 

In \figpanel{tomography}{b}, we show the selected first- to fourth-order moments of the emitted photon field, which account for the thermal population $P_{\mathrm{thermal}}$ in the normalization of the data (see Supplemental Material~\cite{Sup} Sec.~\RNum{4} for details). The solid curves show the expected moment values accounting for thermal excitation \(P_{\mathrm{thermal}}\): the red curve, representing \(\langle \hat{a}_A^\dagger \hat{a}_A \rangle\), is given by \((1-2P_{\mathrm{thermal}})\cos^2(\theta/2)+P_{\mathrm{thermal}}\); the purple curve, corresponding to \(\langle \hat{a}_S^\dagger \hat{a}_S \rangle\), is \((1-2P_{\mathrm{thermal}})\sin^2(\theta/2)+P_{\mathrm{thermal}}\); and the orange curve, representing \(\langle \hat{a}_A^\dagger \hat{a}_S \rangle\), is \((1-2P_{\mathrm{thermal}})\sin(\theta)/2\).

\begin{table}[h]
\caption[ht]{Frequency-bin encoding as an error-detection protocol for photon loss. If any of the two modes get lost before absorbing, the state of Qutrit D at the receiver’s processor will end up in the second excited state, $\left |f\right \rangle$.}
\label{table fidelities}
\centering
\begin{ruledtabular}
\begin{tabular}{ccc}
        & Transmitted photon state & Qutrit at the receiver \\
\hline
Success & $\alpha\left |1\right \rangle_{\omega_A} \left |0\right \rangle_{\omega_S}+\beta\left |0\right \rangle_{\omega_A}\left |1\right \rangle_{\omega_S}$ & $\alpha\left |g\right \rangle_D+\beta\left |e\right \rangle_D$ \\
Error   & $\left |0\right \rangle_{\omega_A}\left |0\right \rangle_{\omega_S}$ &  $\left |f\right \rangle_D$   
\end{tabular}
\end{ruledtabular}
\label{protocol}
\end{table}

\textit{Appendix C: Frequency-bin-encoded photons for error detection ---}
A distributed quantum computing system necessitates both an emitter and a receiver quantum processor, connected via a communication channel that currently exhibits inherent losses. However, with frequency-bin encoding, if any mode is lost during the transmission or not successfully absorbed by the receiver, a vacuum state is received instead of the desired encoded state. Receiving the vacuum state results in Qutrit D at the receiver ending up in the $\left |f\right \rangle_D$ state, instead of the desired state $\alpha\left |g\right \rangle_D+\beta\left |e\right \rangle_D$ (Table~\ref{protocol}). A quantum nondemolition measurement~\cite{ ofek2016extending, kono2018quantum, PhysRevX.8.021003} that distinguishes between the $\left |f\right \rangle$ state and a subspace comprising the $\left |g\right \rangle$ state and the $\left |e\right \rangle$ state, without measuring within that subspace, can detect whether there has been photon loss, without affecting the quantum state of a successfully transmitted frequency-bin-encoded pair. 

\end{document}


\setcounter{figure}{0} 
\renewcommand{\thefigure}{S\arabic{figure}} 
\renewcommand{\theequation}{S\arabic{equation}}

\newcommand{\bluetext}[1]{\textcolor{blue}{#1}}
\newcommand{\figpanel}[2]{Fig.~\hyperref[#1]{\ref*{#1}(#2)}}
\newcommand{\figpanels}[3]{Fig.~\hyperref[#1]{\ref*{#1}(#2-#3)}}
\newcommand{\figpanelNoPrefix}[2]{\hyperref[#1]{\ref*{#1}(#2)}}
\newcommand{\fullfigpanel}[2]{Figure~\hyperref[#1]{\ref*{#1}(#2)}}
\newcommand{\fullfigpanels}[3]{Fig.~\hyperref[#1]{\ref*{#1}(#2, #3)}}
\newcommand{\RNum}[1]{\uppercase\expandafter{\romannumeral #1\relax}}

\author{
Jiaying Yang$^{1,2}$,
Maryam Khanahmadi$^{1}$,
Ingrid Strandberg$^{1}$,
Akshay Gaikwad$^{1}$, 
Claudia Castillo-Moreno$^{1}$,
Anton Frisk Kockum$^{1}$,
Muhammad Asad Ullah$^{2}$,
Göran Johansson$^{1}$,
Axel Martin Eriksson$^{1}$,
Simone Gasparinetti
}

\affiliation{Department of Microtechnology and Nanoscience, Chalmers University of Technology, SE-412 96, G\"{o}teborg, Sweden
\\$^2$  Ericsson Research, Ericsson AB, SE-164 83, Stockholm, Sweden
}

\title{Deterministic generation of frequency-bin-encoded microwave photons --- Supplementary Material} 
\maketitle

\section{A conceptually simple way for frequency-bin encoding}

\begin{figure*}[ht]
\centering
\includegraphics[width=0.35 \linewidth]{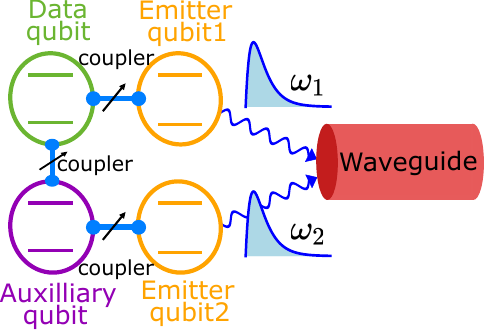}
\caption{A conceptually simple solution for generating frequency-bin encoded photons, which needs two parametric couplers and two emitter qubits coupled to the waveguide. In contrast, our setup, depicted in Fig. 1 in the main text, only requires one data qubit, one coupler, and one emitter qubit.}
\label{frequency-bin-obvious}
\end{figure*}

To implement frequency-bin encoding for photons, where two photonic modes are emitted simultaneously but at different frequencies, we require two emission states and two concurrent transitions. 
A conceptually simple solution to meet this requirement is a four-qubit structure depicted in Fig.~\ref{frequency-bin-obvious} and in Ref.~\cite{o2024deterministic}. 
Initially, the parametric coupler helps generate an entangled state between the data and auxiliary qubits, yielding the process from the initial state \((\alpha \left | g \right \rangle_D + \beta \left | e \right \rangle_D) \left | g \right \rangle_A\) to \(\alpha \left | e \right \rangle_D \left | g \right \rangle_A + \beta \left | g \right \rangle_D \left | e \right \rangle_A\). The data qubit then transfers its state to emitter qubit 1 via the coupler between them, which emits a photon at the qubit frequency \(\omega_1\). Simultaneously, the auxiliary qubit transfers its state to emitter qubit 2, which emits a photon at a different frequency \(\omega_2\). This results in the simultaneous emission of two photons at distinct frequencies, fulfilling the requirements of frequency-bin encoding. In our work, instead, we employ the qubit-coupler-qubit structure for frequency-bin encoding, reducing the number of required components.

\section{Measurement details}
\label{setup}

\begin{figure*}
\centering
\includegraphics[width=0.9 \linewidth]{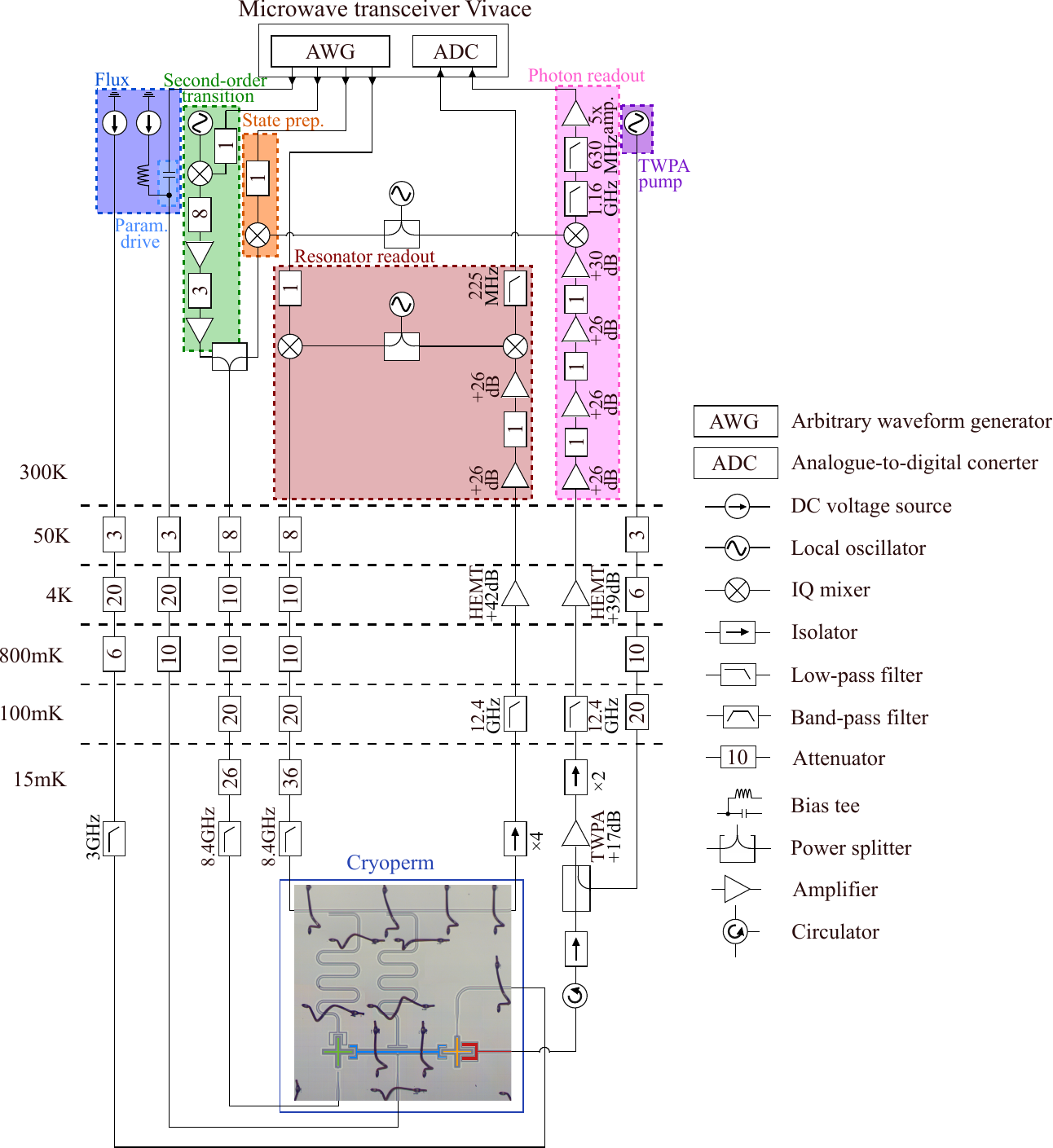}
\caption{The wiring diagram illustrating the connection of the superconducting chip to the input and output lines of the experimental setup.  Note that both the AWG and ADC have two ports for in-phase (I) and quadrature (Q) components; however, only one port is shown for simplicity. The Vivace microwave transceiver board~\cite{Vivace} provides the AWG and ADC channels. Additionally, wire-bonded airbridges (black) are used to ensure a continuous and consistent ground connection, as shown in the device photo.}
\label{wiring diagram}
\end{figure*}

In this work, we utilize a $6.6 \times \SI{6.6}{mm^2}$ superconducting device. The device is fabricated on a silicon substrate, and its RF lines and ground plane consist of aluminium layers deposited on top of the substrate. We wire-bond the device in a copper sample holder, which is then enclosed within a copper shield. To provide additional protection against magnetic interference, we place the bonded device inside another $\mu$-metal shield (cryoperm). This shielding is installed in the mixing chamber of a dilution refrigerator to ensure that all experimental measurements are conducted at temperatures below 15 mK (Fig.~\ref{wiring diagram}).

The device is controlled with one transmission line, one charge line, two flux lines, and one reflection input and output line at the waveguide. The two $\lambda/4$ resonators are coupled to the transmission line to read out the state of Qutrit D and the coupler. The charge line, capacitively coupled to Qutrit D, is used both for preparing the desired quantum state for transfer and for applying the second-order-transition drive. The flux lines, which are inductively coupled to the coupler and the SQUID loops of qubit E, enable frequency tunability. Additionally, we use the flux line connected to the coupler to apply the parametric drive. The waveguide capacitively coupled to Qubit E connects to the reflection input and output lines on the other side, from where the emitted photon field is measured via a traveling-wave parametric amplifier (TWPA)~\cite{macklin2015near}, a high-electron-mobility transistor (HEMT) amplifier, and room-temperature amplifiers in the output line. In our system, we measure the quantum efficiency $\eta$ to be 0.192, where $\eta = \frac{1/2}{1/2+n_{\rm added}}$ and $n_{\rm added}=2.1$ is the added noise photon number by the amplification chain.

In the experiment, data are obtained using a pulsed setup. Microwave control pulses are sent to drive the emitter using arbitrary waveform generators (AWGs), and the data are read out with analogue-to-digital converters (ADCs) from the microwave transceiver platform Vivace~\cite{Vivace}, after up- and down-conversion by IQ mixers and local oscillators. The phase synchronization of the multiple local oscillators and the microwave transceiver is done by connecting them to the same external clock.

The measured parameters of the device are shown in Table~\ref{parameters}, and the coherence times of Qutrit D are shown in Table~\ref{coherent time}. $T_1$ and $T_2$ are the longitudinal and transverse relaxation times of Qutrit D, measured when the frequencies of the coupler and Qubit E are at the sweet spot; $T_1'$ and $T_2'$ are measured when both components are at the operating point. We measure the coherence times for Qutrit D for both the $ge$ and $ef$ transitions. The relaxation rates of the coupler are measured to be $T_1 = 2.3 \pm 0.5 \, \mu\text{s}$ and $T_2 = 1.4 \pm 0.4 \, \mu\text{s}$.

\begin{table}
\caption[ht]{Experimentally measured parameter values.}
\centering
\scalebox{1}{
\begin{tabular}{ccccc}
\toprule
Parameter & Symbol              & Data qutrit & Coupler & Emitter qubit  \\
\midrule
Readout resonator frequency &$f_R/2\pi$ (GHz)       & 6.19    & 6.74 & --          \\
Decay rate of the resonator to the feedline &$\kappa/2\pi$ (kHz)    & 523 & 752 & --          \\
Coupling rate between Qutrit D (coupler) and their readout resonator&$j/2\pi$ (MHz)         & 69.33 & 42.34 & --          \\
Qutrit (coupler, or qubit) frequency without DC flux bias &$\omega/2\pi$ (GHz)    & 5.05    & 8.46   & 6.17       \\
Qutrit (coupler, or qubit) frequency with DC flux bias to operating point&$\omega'/2\pi$ (GHz)   & 5.05    & 5.79   & 5.70        \\
Anharmonicity of qutrit (coupler, or qubit)&$\alpha/2\pi$ (MHz)    & -215    & -60    &  -215 \\
Coupling rate between Qutrit D (Qubit E) and  coupler &$g/2\pi$ (MHz)         &   37.5  & --     & 46      \\
Coupling rate between Qubit E and the coplanar waveguide &$\Gamma_{E}/2\pi$ (MHz)    & --      & --     & 8.0         \\
\bottomrule
\end{tabular}
}
\label{parameters}
\end{table}

\begin{table}[h!]
\caption[ht]{The coherence times of Qutrit D. We have larger $T_{1, ef}$ than $T_{1, ge}$, which can be attributed to the weaker coupling of the system to the environment at $\omega_D^{ef}$ compared to $\omega_D^{ge}$. }
\centering
\scalebox{1}{
\begin{tabular}{cccc|cccc}
\toprule
\multicolumn{4}{c|}{$T_1$ ($\mu$s)} & \multicolumn{4}{c}{$T_2$ ($\mu$s)} \\
\midrule
\multicolumn{2}{c}{Sweet spot} & \multicolumn{2}{c|}{Operating spot} & \multicolumn{2}{c}{Sweet spot} & \multicolumn{2}{c}{Operating spot} \\
\midrule
$T_{1, ge}$ & $T_{1, ef}$ & $T'_{1, ge}$ & $T'_{1, ef}$ & $T_{2, ge}$ & $T_{2, ef}$ & $T'_{2, ge}$ & $T'_{2, ef}$ \\
\midrule
 $21.7\pm3.6$& $27.0\pm3.0$&$20.8\pm2.5$&$27.0\pm3.0$&$5.7\pm0.7$&  $4.4 \pm 0.5$&$14.0 \pm 3.4$&$8.8\pm 1.4$    
\\ \bottomrule
\end{tabular}
\label{coherent time}
}
\end{table}

\newpage
\section{Theoretical model}
\subsection{Bare Hamiltonian}
According to Fig.~1 in the main text,  we consider Qutrit D as a three-level system with frequency $\bm{\omega}_D$ and the anharmonicity $\alpha = 2\chi_d$ and the coupler and Qubit E as two-level systems, since large anharmonicity suppresses populating the higher energy levels. In the following, we consider the annihilation and creation operators $\{(\bm{\hat{d}}, \bm{\hat{d}}^\dagger), (\bm{\hat{c}},\bm{\hat{c}}^\dagger), (\bm{\hat{e}},\bm{\hat{e}}^\dagger)\}$, which correspond to Qutrit D, the coupler, and Qubit E, respectively. Considering the same frequency for the emitter and the coupler as $\bm{\omega_E}$ and assuming the coupling between the coupler and Qutrit D (Qubit E) as $g_{dc}$ $(g_{ec})$, the total Hamiltonian in the lab frame is obtained as
\begin{equation}
\begin{aligned}
    H =& \bm{\omega_D \hat{d}^\dagger \hat{d}} + \bm{\omega_E \hat{e}^\dagger \hat{e}}+ (\bm{\omega_E} +\bm{\omega_{\mathrm{{ext}}}}(t)) \bm{\hat{c}^\dagger \hat{c}} + \chi_{d}\bm{\hat{d}^{\dagger 2} \hat{d}^2}+ g_{dc}\bm{(\hat{d}+\hat{d}^\dagger)(\hat{c}+\hat{c}^\dagger)}+ g_{ec}\bm{(\hat{e}+\hat{e}^\dagger)(\hat{c}+\hat{c}^\dagger)} \\
    &+ \zeta' (e^{i(2\omega_d +2\chi_d - (\omega_e +\delta))}\bm{\hat{d}}+e^{-i(2\omega_d +2\chi_d - (\omega_e +\delta))}\bm{\hat{d}}^\dagger)\, ,
\end{aligned}
\end{equation}
where $\bm{\omega_{\mathrm{ext}}}(t)$ is the time-dependent frequency corresponding to the flux drive of the coupler and the amplitude of drive on Qutrit D is considered to be $\zeta'$. Note that $g_{ec}$ is written as $g$ in the main text for simplicity. Introducing two hybridized modes, symmetric $\bm{\hat{a}_S}= (\bm{\hat{e}}+\bm{\hat{c}})/\sqrt{2}$ and antisymmetric $\bm{\hat{a}_A} = (\bm{\hat{e}}-\bm{\hat{c}})/\sqrt{2}$, respectively, the Hamiltonian changes to
\begin{equation}
\begin{aligned}
   H =&  \omega_D \bm{\hat{d}^\dagger \hat{d}} + \bm{\omega_S} \bm{\hat{a}_S}^\dagger \bm{\hat{a}_S} +\bm{\omega_A} \bm{\hat{a}_A}^\dagger \bm{\hat{a}_A} +
   \frac{g_{dc}}{\sqrt{2}}[\bm{\hat{d}}^\dagger(\bm{\hat{a}_S}-\bm{\hat{a}_A})+(\bm{\hat{a}_S}-\bm{\hat{a}_A})^\dagger \bm{\hat{d}}]\\
   & +\frac{\bm{\omega_{\mathrm{ext}}}(t)}{2} (\bm{\hat{a}_S}-\bm{\hat{a}_A})^\dagger (\bm{\hat{a}_S}-\bm{\hat{a}_A}) + \chi_{d}\bm{\hat{d}}^{\dagger 2} \bm{\hat{d}}^2 + \zeta' (e^{i(2\omega_d+2\chi_d - \omega_S)}\bm{\hat{d}}+e^{-i(2\omega_d +2\chi_d - \omega_S)}\bm{\hat{d}}^\dagger)\, ,
\end{aligned} \label{barHam}
\end{equation}
where the hybridized frequencies are $\omega_S= (\bm{\omega_E}+g_{ec})$ and $\omega_A= (\bm{\omega_E}-g_{ec})$. The  frequency of the coupler is evaluated as
\begin{align}
    \bm{\omega_C(t)} = \omega_c^0 \sqrt{|\cos(\frac{\pi \phi(t)}{\phi_0})|} \equiv\bm{\omega_{E}} + \bm{\omega_{\mathrm{ext}}(t)} \rightarrow  \pi\phi(t)/\phi_0 =\varphi_{dc} + \eta' \cos(\omega_{d} t)\, ,
\end{align}
where in a weak modulation drive, the Taylor expansion of the coupler frequency leads to
\begin{align}
\bm{\omega_E(t)} =& \, \omega_c^0 \sqrt{|\cos( \varphi_{dc})|}\, ,\\
    \bm{\omega_{\mathrm{ext}}(t)} = 
    &\,\frac{\partial \bm{\omega_C}}{\partial \varphi}|_{\varphi=\varphi_{dc}} \eta' \cos((\bm{\omega_E}-\bm{\omega_D})t) + \frac{1}{2}\frac{\partial^2 \bm{\omega_C}}{\partial \varphi^2}|_{\varphi=\varphi_{dc}} \eta^{'2} \cos^2((\bm{\omega_E}-\bm{\omega_D})t)\, .
\end{align}
The emitter qubit is strongly coupled to a waveguide with strength $L = \sqrt{\Gamma_E} \bm{\hat{e}} $,  which in the symmetric and antisymmetric basis changes to $L = \sqrt{\Gamma_E} \frac{\bm{\hat{a}_S}+\bm{\hat{a}_A}}{\sqrt{2}}$. Applying the secular approximation in the rotating frame of $\bm{\omega}_S$ and $\bm{\omega}_A$, leads to two separate decay channels $L_1 =\sqrt{\frac{\Gamma_E}{2}} \bm{\hat{a}_S}, L_2=\sqrt{\frac{\Gamma_E}{2}} \bm{\hat{a}_A}$ with half the decay rate of the original emitter coupling.


\subsection{Effective Hamiltonian}
Utilizing the black-box quantization methods~\cite{nigg2012black}, one can find the dressed-mode coefficients from the linear and time-independent parts of the Hamiltonian; the first line of Eq. \eqref{barHam}. Introducing the dressed modes $\bm{\hat{d}} = \varphi_d \hat{d}+\varphi_A \hat{a}_A+\varphi_S \hat{a}_S$, $ \bm{\hat{a}}_S = \varphi'_d \hat{d}+\varphi'_A \hat{a}_A+\varphi'_S \hat{a}_S$, and $\bm{\hat{a}}_A = \varphi''_d \hat{d}+\varphi''_A \hat{a}_A+\varphi''_S \hat{a}_S$, diagonalizing the first line of Eq.~\eqref{barHam}, the second line can be written in the basis $\hat{d},\hat{a}_A,\hat{a}_S$. Applying the optimal displacement transformation $\alpha = \frac{-\zeta}{\omega_d +2\chi_d - \omega_S} \equiv \epsilon$ on the $d$ mode, in the rotating frame of $\omega_D \hat{d}^\dagger \hat{d} + \omega_S \hat{a}_S^\dagger \hat{a}_S +\omega_A \hat{a}_A^\dagger \hat{a}_A $, the effective Hamiltonian is obtained as
\begin{align}\label{heff}
H_{\mathrm{eff}} = & \,\eta(\hat{d}\hat{a}_A^\dagger+\hat{a}_A \hat{d}^\dagger) + 2\epsilon \chi_d \varphi_{d}^3\varphi_{S}[  e^{-i2\chi_d t}\hat{d}^{\dagger 2}\hat{a}_S + e^{i2\chi_dt}\hat{a}_S^\dagger \hat{d}^2]\\
& + \chi_{d}\varphi_{d}^4 \hat{d}^{\dagger 2} \hat{d}^2+  4 \chi_{d}\varphi_{d}^2\varphi_{S}^2 \hat{d}^\dagger \hat{d} \hat{a}_S^\dagger \hat{a}_S+4\chi_{d}\varphi_{d}^2\varphi_{A}^2 \hat{d}^\dagger \hat{d} \hat{a}_A^\dagger \hat{a}_A\nonumber\\
&+ 4 \epsilon^2\chi_{d}\varphi_{d}^4 \hat{d}^{\dagger } \hat{d} +4 \epsilon^2 \chi_{d}\varphi_{d}^2\varphi_{S}^2 \hat{a}_S^{\dagger } \hat{a}_S +4 \epsilon^2 \chi_{d}\varphi_{d}^2\varphi_{A}^2 \hat{a}_A^{\dagger } \hat{a}_A\nonumber\\
& + S(\eta(t))\bigg[(\varphi'_d-\varphi''_d)^2 \hat{d}^{\dagger } \hat{d} +(\varphi'_{S}-\varphi''_S)^2 \hat{a}_S^{\dagger } \hat{a}_S +(\varphi'_{A}-\varphi''_A)^2 \hat{a}_A^{\dagger } \hat{a}_A \bigg],\nonumber
\end{align}
where the amplitude of the parametric and the second-order-transition drive mentioned in the main text Eq.~(1) corresponds to 
\begin{align}\label{amp}
    \eta = \frac{1}{2}\frac{\partial \omega_c}{\partial \phi}|_{\phi=\phi_{dc}} \eta' (\varphi'_d-\varphi''_d) (\varphi'_{A}-\varphi''_{A})\, ,\, \zeta = 2\epsilon \chi_d \varphi_{d}^3\varphi_{s}e^{-i2\chi_dt}\, ,
\end{align}
respectively, and the Stark shift induced by the parametric drive is obtained as
\begin{align}
     S(\eta') = \frac{1}{4}\frac{\partial^2 \omega_c}{\partial \varphi^2}|_{\varphi=\varphi_{dc}} \eta^{'2}.
\end{align}
It is important to note that the Stark shift obtained from the second-order-transition drive on Qutrit D is comparable to the detuning $\omega_d-\omega_S$. To avoid slowing down the release process, we consider the Stark shift in the frequency of both the second-order-transition and parametric drives. Introducing the AC Stark shift on Qutrit D, $\omega_{AC} = 4 \epsilon^2\chi_{d}\varphi_{d}^4$, the frequencies of the parametric and second-order-transition drives change to $\omega_{\mathrm{param}} = \omega_d - \omega_A +\omega_{AC}$ and $\omega_{\rm 2nd} = 2\omega_d +2\chi_d - \omega_S + 2\omega_{AC} $, respectively, where the amplitudes in Eq.~\eqref{amp} change to $\eta \rightarrow \eta e^{-i\omega_{AC}t}\,\,,\,\,\zeta \rightarrow \zeta e^{-i2\omega_{AC}t}$. 

It is worth noting that the second-order-transition and the parametric drives, by applying the mode decomposition method mentioned in \cite{PhysRevResearch.5.043071}, are optimized to populate one single mode (wave packet), in each decay channel. To find the mode decomposition of the output fields, we calculate the two-time-correlation function for each channel, symmetric and antisymmetric, with indices S and A, respectively: 
\begin{align}\label{g2}
G^{(2)}_{S(A)}(t_1,t_2) = \langle L_{S(A)}^\dagger(t_1) L_{S(A)}(t_2)\rangle \equiv n_{S(A)} f_{S(A)}^*(t_2)f_{S(A)}(t_1) ,   
\end{align}
where the number of photons per mode is $n_{S(A)}$ with the corresponding shape $f_{S(A)}(t)$ utilized in Sec.~\ref{charact}-A. It is worth noting that in Eq.~(\ref{g2}) we consider two separate Lindblad operators to find the temporal mode's shape, corresponding to symmetric and antisymmetric propagating wavepackets, since the overlap between two modes is on the order of $10^{-4}$. However in the general case, one can utilize the Lindblad operator $L = \sqrt{\Gamma_E/2}(e^{i\omega_S t}\hat{a}_S + e^{i\omega_A t}\hat{a}_A)$ and by the mode decomposition $G(t_1, t_2) =\left \langle  L_A(t_1)^\dagger L_S(t_2) \right \rangle = n_A f^*_A(t_2)f_A(t_1)+ n_S f^*_S(t_2)f_S(t_1)$ find the temporal shape of the propagating modes. Note that we neglect the population of other unwanted temporal modes as the output field highly populates one wavepacket for each mode $A$ and $S$.
The two-time correlation function is a Hermitian operator where the eigenvalue is a positive real number $n_A,n_S$ and the eigenmodes are orthonormal, together leading to $\int \text{d}t f^*_A(t)f_S(t) =0$. According to \cite{penas2024multiplexed}, when the two frequency modes are detuned by $(\omega_A -\omega_S)/(\Gamma_E/2)$ = 23, the reconstructed state exhibits an infidelity on the order of $10^{-5}$, indicating that the temporal overlap of the frequency modes has a minimal impact on fidelity.

To simulate the full tomography on the output field, we consider two catching cavities described by the annihilation operators $\hat{v}_S,\hat{v}_A$ to catch each traveling wave packet $f_S(t),f_A(t)$, respectively \cite{PhysRevLett.123.123604}. This can be achieved by considering the cascaded formalism \cite{PhysRevLett.70.2273} between the sender, including Qutrit D, the symmetric, and the antisymmetric modes, and the two catching cavities by introducing the total Lindblad operator $L_{S(A)} = \sqrt{\frac{\gamma}{2}} \hat{a}_{S(A)} + g_{S(A)}^*(t) \hat{v}_{S(A)}$, where the time-dependent decay rate obtained as $g_{S(A)}(t) = \frac{-f^*_{S(A)}(t)}{\sqrt{\int_0^t \mathrm{d}t' |f_{S(A)}|^2}}$. Considering the cascaded Hamiltonian $H_{\mathrm{total}} = H_{\mathrm{eff}} + \frac{i\sqrt{\gamma}}{2\sqrt{2}} \big[  g_S^*(t) \hat{a}_S^\dagger \hat{v}_S - g_S(t) \hat{a}_S \hat{v}_S^\dagger\big] + \frac{i\sqrt{\gamma}}{2\sqrt{2}} \big[  g_A^*(t) \hat{a}_A^\dagger \hat{v}_A - g_A(t) \hat{a}_A \hat{v}_A^\dagger\big] $ and the total Lindblad operators $L_S, L_A$, the final state of the corresponding Lindblad master equation provides the density matrix of the two-mode traveling quantum state, where the sender transitions to the vacuum state and decouples from the traveling wavepacket.



\section{Characterization of the frequency-encoded photonic modes}\label{charact}
\subsection{Temporal mode matching}

We apply two temporal filters $f_k(t)$ ($k=A,S$) to the output field, extracting the frequency-bin encoded propagating modes $\hat{a}_k$ out of the time-dependent output field $\hat{a}^{\rm out} (t)$ as~\cite{LoudonQuantumTheory2000}
\begin{equation}
    \label{eq_TM}
        \hat{a}_k =\int_{-\infty}^{\infty} dt f_k(t) \hat a_{\text{out}}(t),
\end{equation}
where $\hat a_{\text{out}}(t)$ is given by the input-output relation~\cite{LoudonQuantumTheory2000,GardinerQuantumNoise2004}, $ \hat a_{\text{out}}(t)=\sqrt{\Gamma}\hat \sigma(t)-\hat a_{\text{in}}(t)$, with $\hat a_{\text{in}}(t)$ the input field. $f_k(t)$ ($k=A,S$) are the two temporal filters with carrier frequencies $\omega_A$ and $\omega_S$ that fulfil
\begin{equation}\label{eq:temporal_filter}
f_k(t)=v_k(t)\cdot e^{i(\omega_k t+\phi)}\, . 
\end{equation}
Here, $v_k(t)$ represents the wavepacket profile for both filters, which has the same shape as the emitted photon envelopes [Fig.~3(c) in the main text] to enable the highest matching efficiency. 
The extracted modes fulfil bosonic commutation relations $[\hat a_k,\hat  a_k^\dagger]=1$, ensured by the normalization condition for the filter function $\int_{-\infty}^\infty dt |f_k(t)|^2=1$. 

\subsection{Denoising of moments}
Due to measurement noise arising from cable losses, amplification chain, mode-matching inefficiency, and other factors, the directly obtained total mode from temporal template matching, denoted as $\hat{S}_k$ $(k=A, S)$, does not solely represent the target modes $\hat{a}^{\rm out}_k$. Instead, $\hat{S}_k$ comprises both the target mode $\hat{a}_k^{\rm out}$ and an additional noise mode $\hat{h}^\dagger$, $\hat{S}_k = \hat{a}_k^{\rm out} + \hat{h}^\dagger$. To remove the noise mode $\hat{h}^\dagger$, we operate an interleaved measurement, to sweep between two cases with and without both the parametric and second-order-transition drives. In the first case, we measure the total mode including both the targeted mode and the noise mode. In the second case, the target mode is left in vacuum and the measurement can be served as a reference of the noise mode. The switching between the two cases is repeated $n$ = $5\times 10^6$ times. We then calculate the averaged moments from these repetitions. 

The first- and second-order moments of the two propagating modes, $\left<\hat{a}_k^{\rm out}\right>$ and $\left<(\hat{a}_{k}^{\rm out} )^\dagger\hat{a}_{k}^{\rm out}\right>$, are obtained by~\cite{eichlor_thesis}
\begin{equation}
\begin{split}
\left<\hat{a}_k^{\rm out}\right> &= \left<\hat{S}_k-\hat{h}^\dagger\right>\, , \\
\left<(\hat{a}_{k}^{\rm out} )^\dagger\hat{a}_{k}^{\rm out}\right> &= \left<\hat{S}_{k}^\dagger\hat{S}_{k}-\hat{S}_{k}\hat{h}-\hat{S}_{k}^\dagger\hat{h}^\dagger+\hat{h}\hat{h}^\dagger\right>\, ,
\end{split}
\label{moments_cal_2D}
\end{equation}
where the angle brackets represent the averaging over the $n$ repetitions. Generalized to fourth-order moments or more, the moments are computed as follows~\cite{eichlor_thesis}:

\begin{widetext}
\begin{multline}
\langle(\hat{S_A}^{\dagger})^{m_A} \hat{S_A}^{n_A} (\hat{S_S}^{\dagger})^{m_S} \hat{S_S}^{n_S}\rangle=\sum_{i_A, j_A, i_S,j_S =0}^{m_A, n_A, m_S, n_S}\left(\begin{array}{c}{m_A} \\ {i_A}\end{array}\right)\left(\begin{array}{c}{n_A} \\ {j_A}\end{array}\right) \left(\begin{array}{c}{m_S} \\ {i_S}\end{array}\right) \left(\begin{array}{c}{n_S} \\ {j_S}\end{array}\right) \\
\times \langle (\hat{a}_A^{\dagger})^{i_A} \hat{a}_A^{j_A} (\hat{a}_S^{\dagger})^{i_S} \hat{a}_S^{j_S}\rangle \langle(\hat{h}_A^{\dagger})^{m_A-i_A} \hat{h}_A^{n_A-j_A}(\hat{h}_S^{\dagger})^{m_S-i_S}\hat{h}_S^{n_S-j_S}\rangle\, .
\label{cal_moments}
\end{multline}
\end{widetext}

\subsection{Normalization of moments}

\begin{figure*}[ht]
\centering
\includegraphics[width=0.45 \linewidth]{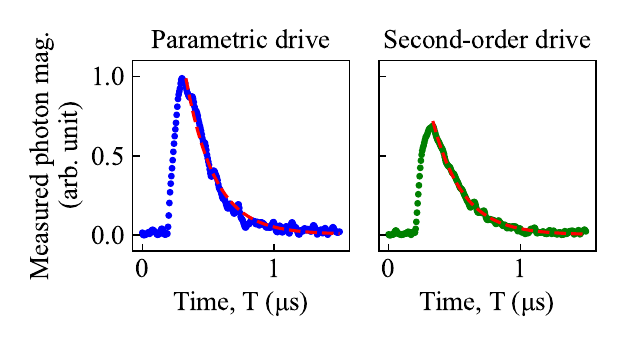}
\caption{Fitting of the emitter photon to obtain effective decay rate. The filled circles represent the measured photon magnitude, showing the same data as Fig.~3(c) in the main text. The dashed curves are the exponential fit.}
\label{eff decay}
\end{figure*}

The moments are normalized according to the assumption that, when the parametric drive or the second-order transition is turned on, the decay of Qutrit D occurs via two independent channels: the detected modes of the waveguide, and the intrinsic losses of Qutrit D, with rates $\Gamma_{\rm eff}$ and $\Gamma_{\rm D}$, respectively. Accordingly, when Qutrit D is prepared in state $|e\rangle$ or $|f\rangle$ for photon emission in frequency $\omega_A$ and $\omega_S$, respectively, we assume that the second-order moments are $\langle \hat a_A^{\dagger} \hat a_A \rangle=\Gamma_{\rm eff}^A/(\Gamma_{\rm eff}^A+\Gamma_{\rm D}^{ge})(1-P_{\mathrm{thermal}})$ and $\langle \hat a_S^{\dagger} \hat a_S \rangle=\Gamma_{\rm eff}^S/(\Gamma_{\rm eff}^S+\Gamma_{\rm D}^{ef})(1-P_{\mathrm{thermal}})$, where $P_{\mathrm{thermal}}$ is the thermally exited population of Qutrit D. Across our datasets, $P_{\mathrm{thermal}}$ ranges between 13.4\% and 16.0\% determined by solving $\langle \hat a_A^{\dagger} \hat a_A \rangle+\langle \hat a_S^{\dagger} \hat a_S \rangle = \Gamma_{\rm eff}^A/(\Gamma_{\rm eff}^A+\Gamma_{\rm D}^{ge})$ for Qutrit D prepared in state $|e\rangle$. $\Gamma_{\rm eff}^A/2\pi = \SI{4.59}{MHz}$ and $\Gamma_{\rm eff}^S/2\pi = \SI{4.39}{MHz}$ are the effective decay rate of the propagating modes obtained by exponentially fitting the emitted photon envelopes (Fig.~\ref{eff decay}). $\Gamma_{\rm D}^{ge}$ and $\Gamma_{\rm D}^{ef}$ are longitudinal relaxation rates of Qutrit D, calculated as $\Gamma_{\rm D}^{ge} = 1/T'_{1, ge}$ and $\Gamma_{\rm D}^{ef} = 1/  T'_{1, ef}$.

\subsection{Joint quantum state tomography and process tomography}
\subsubsection{Least-squares method}
In the joint quantum state tomography (QST) of the two propagating modes, we reconstruct the optimal density matrix from the first- to fourth-order moments of the two modes using least-squares (LS) optimization~\cite{PhysRevResearch.2.042002, PhysRevApplied.18.044041}. Mathematically, the LS method solves the following convex
optimization problem to find the optimal density matrix $\rho$: 
\begin{subequations}
\begin{alignat}{2}
&\!\min_{\rho}        &\qquad& \left\Vert (\overrightarrow{\mathcal{B}}-\mathcal{A} \overrightarrow{\rho})\right\Vert_{\ell_2} \label{obj_ls}\, ,\\
& \text{subject to}   &      & \rho \geq 0\, ,\label{obj:constraint1_ls}\\
&                     &      & \text{Tr}(\rho) = 1 \, ,\label{obj:constraint2_ls}
\end{alignat}
\end{subequations}
where $\overrightarrow{\rho}$ in the objective function given in Eq.~(\ref{obj_ls}) represents the vectorized form of the density matrix.
In Eq.~(\ref{obj_ls}),  $\overrightarrow{\mathcal{B}}$  is a column vector containing the experimentally measured first- to fourth-order moments, $\left \langle (\hat{a}_A^{\dagger})^{m_A} \hat{a}_A^{n_A} (\hat{a}_S^{\dagger})^{m_S} \hat{a}_S^{n_S} \right \rangle $, for $m_A, n_A, m_S, n_S \in \left \{ 0, 1, 2 \right \} $. The matrix $\mathcal{A}$, commonly known as the sensing matrix ~\cite{PhysRevApplied.18.044041}, is determined solely by the operator basis set ($\{ \vert i \rangle \langle j \vert \}$)  
and the measurement observable set ($\{ (\hat{a}_A^{\dagger})^{m_A} \hat{a}_A^{n_A} (\hat{a}_S^{\dagger})^{m_S} \hat{a}_S^{n_S} \}$). Furthermore, \( \|\cdot\|_{\ell_2} \), representing the \(\ell_2\) norm (also known as the Euclidean norm), is calculated as the square root of the sum of the squared components of a vector.
The additional constraints given in  Eq.~(\ref{obj:constraint1_ls}-\ref{obj:constraint2_ls}) are completely positive (CP) and trace-preserving (TP) conditions of the density matrix, ensuring its physical validity. The convex optimization problem is solved using CVXPY~\cite{diamond2016cvxpy}. 

To carry out quantum process tomography (QPT) of the state transfer scheme, we initialize Qutrit D into four of the cardinal states,  $\left |g \right \rangle_D$, $\frac{1}{\sqrt{2}} (\left |g \right \rangle_D + \left |e \right \rangle_D)$, $\left |e \right \rangle_D$, and $\frac{1}{\sqrt{2}} (\left |g \right \rangle_D -i \left |e \right \rangle_D)$. After the parametric drive and the second-order-transition drive, the state of Qutrit D is transferred to the frequency-bin-encoded photonic modes. We then perform QST on the propagating modes followed by projection onto the space spanned by $\left |1\right \rangle_{\omega_A} \left |0\right \rangle_{\omega_S}$ and $\left |0\right \rangle_{\omega_A}\left |1\right \rangle_{\omega_S}$  for each initial state. Here we use the Kraus-operator formalism to represent the quantum process as~\cite{gaikwad-sr-2022}
\begin{equation} \label{qpt}
    \rho_{out} = \Lambda(\rho_{in})= \sum_{m,n = 0}^3\chi_{mn}\sigma_m\rho_{in}\sigma_n^\dagger\, ,
\end{equation}
where $\Lambda(\cdot)$ represents the quantum map; in our case, it is the state-transfer process from Qutrit D to frequency-bin-encoded photonic modes. The quantities $\chi_{mn}$ are elements of the process matrix $\chi$ characterizing $\Lambda$. The $\{ \sigma_i \}$ are a fixed set of basis operators which are set to be Pauli operators, $\{{\sigma}_0, {\sigma}_1, {\sigma}_2, {\sigma}_3\} = \{{I}, {\sigma}_x, {\sigma}_y, {\sigma}_z\}$.
Using Eq.~(\ref{qpt}) with the four cardinal states (the expected input probe states considering thermal excitation)
and corresponding frequency-bin-encoded photon states (output states from quantum state tomography) we can form a system of linear equations of the form $\mathcal{M} \Vec{\chi} = \mathcal{\Vec{D}}$, where $\mathcal{\Vec{D}}$ is constructed from $\rho_{out}$ and $\Vec{\chi}$ represents the vectorized form of the $\chi$ matrix. The coefficient matrix $\mathcal{M}$ [similar to the sensing matrix $\mathcal{A}$ in Eq.~(\ref{obj_ls})] depends only on the set of probe states and the measurement observables; in our case, the cardinal states and the moment operators. The recipe to reconstruct $\mathcal{M}$ can be found in Ref.~\cite{ak-qip-2022}. To calculate a physically valid $\chi$ matrix characterizing an underlying CPTP map, we form a LS optimization problem (similar to QST) as follows,
\begin{subequations}
\begin{alignat}{2}
&\!\min_{\chi}        &\qquad& \left\Vert (\overrightarrow{\mathcal{D}}-\mathcal{M} \overrightarrow{\chi})\right\Vert_{\ell_2} \label{obj_qpt}\, ,\\
& \text{subject to}   &      & \chi \geq 0\, ,\label{obj:constraint1_qpt}\\
&                     &      & \sum_{m,n = 0}^3\chi_{mn} \sigma_m\sigma_n= I \, ,\label{obj:constraint2_qpt}
\end{alignat}
\end{subequations}
where Eq.~(\ref{obj:constraint1_qpt}) and Eq.~(\ref{obj:constraint2_qpt}) ensure the CP and TP conditions respectively. Again, we use the  CVXPY~\cite{diamond2016cvxpy} tool to solve the convex optimization problem. The process fidelity is defined as $F(\chi_{\rm ideal}, \chi) = \left(\mathrm{Tr}\sqrt{\sqrt{\chi_{\rm ideal}}\chi\sqrt{\chi_{\rm ideal}}}\right)^2$, with $\chi_{\rm ideal}$ the process matrix representing the ideal unit process.

\subsubsection{Detailed tomography results}

\begin{figure*}[ht]
\centering
\includegraphics[width=0.6\linewidth]{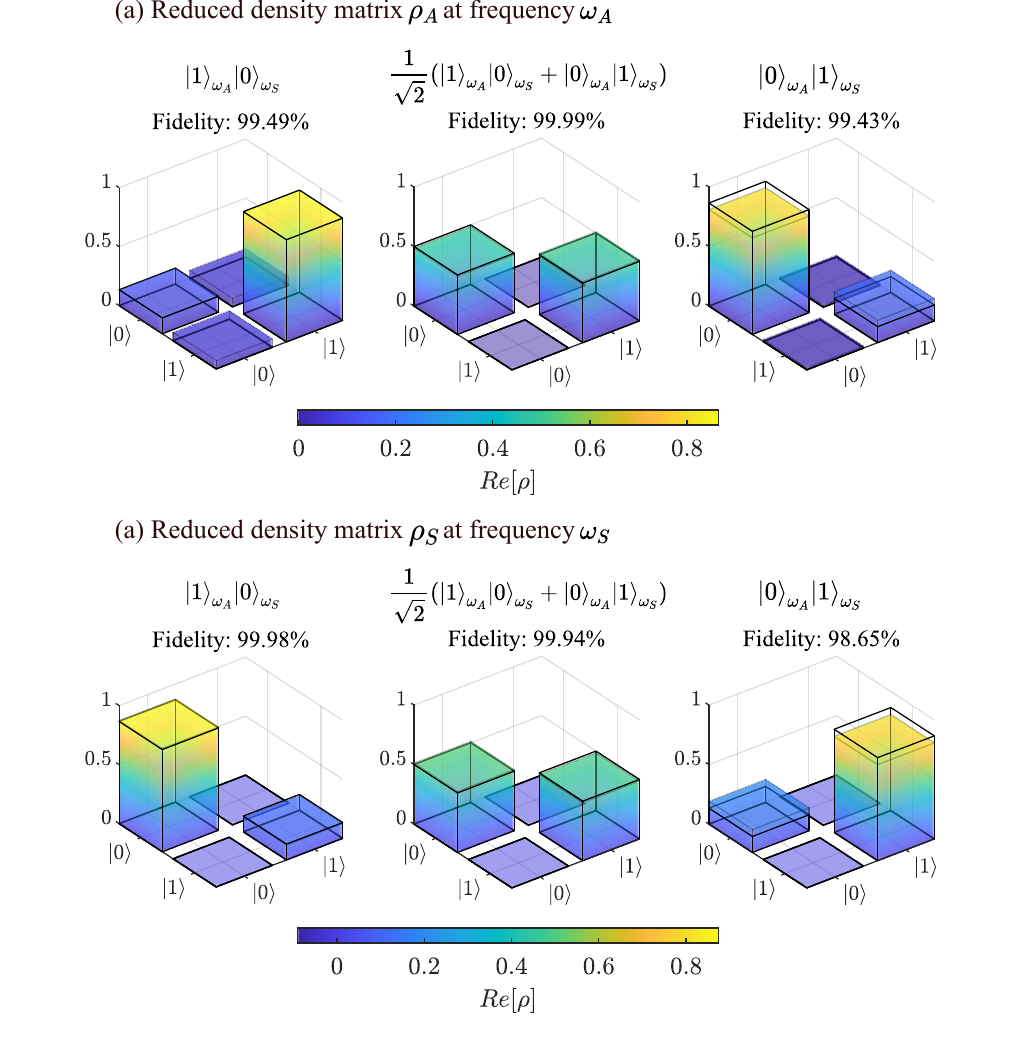}
\caption{The reconstructed (colored bars) and expected (black frames) reduced density matrices of each frequency-bin-encoded photonic mode are shown, obtained by taking the partial trace of the two-mode density matrix, leaving only the single-mode density matrices at frequencies (a) $\omega_A$ and (b) $\omega_S$, respectively. We present these matrices for three different initial states of Qutrit D: $\left |g  \right \rangle_D$, $\frac{1}{\sqrt{2}} (\left |g  \right \rangle_D+\left |e  \right \rangle_D)$, and $\left |e  \right \rangle_D$. The fidelity $F$ between the reconstructed and expected matrices (accounting for thermal population) is indicated above each corresponding matrix.}
\label{tomography_details}
\end{figure*}

From joint QST based on LS optimization, we obtain the density matrices of the two frequency-bin-encoded photonic modes, shown in Fig.~4(c-e) of the main text. Here, by tracing out one of the photonic states, we are left with the reduced density matrices of two single modes in frequencies $\omega_A$ and $\omega_S$, respectively [\fullfigpanels{tomography_details}{a}{b}].

In quantum process tomography, the four cardinal states used for reconstructing the process matrix $\chi$, along with their corresponding fidelities, are presented in Table~\ref{process tomography}. The second column lists the fidelities of the density matrices obtained from LS optimization, computed with respect to the thermally populated initial states. The first three states are depicted in the joint QST results [Fig.~4(c-e) in the main text], while the fourth state is an additional one necessary for QPT. In Table~\ref{process tomography}, we report two fidelity measures, defined as \(
F(\rho_{\bullet}, \rho) 
= 
\left(\mathrm{Tr} \sqrt{\sqrt{\rho_{\bullet}} \, \rho \, \sqrt{\rho_{\bullet}}} \right)^{2}
\), where \(\rho_{\bullet}\) can represent either a thermal state (\(\rho_{\rm in}\)) or the ideal state (\(\rho_{\rm ideal}\)). 
The thermal state \(\rho_{\rm in}\) is obtained by taking into account the initial thermal population of Qutrit D, as shown in the main text. Although a more refined calculation would include thermal noise in both the coupler and Qubit E, here we provide a stronger comparison by listing the fidelities with respect to the ideal state along with that to the thermal state---both experimentally and in simulation---in Table~\ref{process tomography}.

\begin{table}[h!]
\caption[ht]{Fidelity of reconstructed density matrices for the four cardinal states used in quantum process tomography. Experimental fidelities are obtained from LS-QST and compared with numerical simulations. In both cases, fidelities are evaluated with respect to density matrices that either include or exclude the initial thermal population effects.}
\centering

\scalebox{1}{
\begin{tabular}{l|cc|cc}
\toprule
& \multicolumn{2}{c|}{Measured fidelity from LS-QST} 
& \multicolumn{2}{c}{Simulated fidelity} 
\\
\cmidrule(lr){2-3}
\cmidrule(lr){4-5}
State 
& w.r.t. \(\rho_{\rm in}\) 
& w.r.t. \(\rho_{\rm ideal}\)  
& w.r.t. \(\rho_{\rm in}\)   
& w.r.t. \(\rho_{\rm ideal}\)\\
\midrule
$\left |1\right \rangle_{\omega_A} \left |0\right \rangle_{\omega_S}$ 
& \SI{99.02}{\percent}  
& \SI{86.43}{\percent}  
& \SI{99.24}{\percent}
& \SI{86.53}{\percent}\\

$(\left |1\right \rangle_{\omega_A} \left |0\right \rangle_{\omega_S}+\left |0\right \rangle_{\omega_A}\left |1\right \rangle_{\omega_A})/\sqrt{2}$
& \SI{98.51}{\percent}  
& \SI{83.58}{\percent}
& \SI{98.97}{\percent}
& \SI{85.66}{\percent}\\

$\left |0\right \rangle_{\omega_A}\left |1\right \rangle_{\omega_S}$
& \SI{94.57}{\percent}
& \SI{78.54}{\percent}
& \SI{98.68}{\percent}
& \SI{85.40}{\percent}\\

$(\left |1\right \rangle_{\omega_A} \left |0\right \rangle_{\omega_S}
-i\left |0\right \rangle_{\omega_A}\left |1\right \rangle_{\omega_A})/\sqrt{2}$
& \SI{98.04}{\percent}
& \SI{81.17}{\percent}
& \SI{98.97}{\percent}
& \SI{85.66}{\percent}\\
\midrule
Average 
& \SI{97.53}{\percent}
& \SI{82.43}{\percent}
& \SI{98.97}{\percent}
& \SI{85.81}{\percent}\\
\bottomrule
\end{tabular}
}
\label{process tomography}
\end{table}

\section{Protocol of quantum state transfer and remote entanglement}

\begin{figure*}[ht]
\centering
\includegraphics[width=0.92\linewidth]{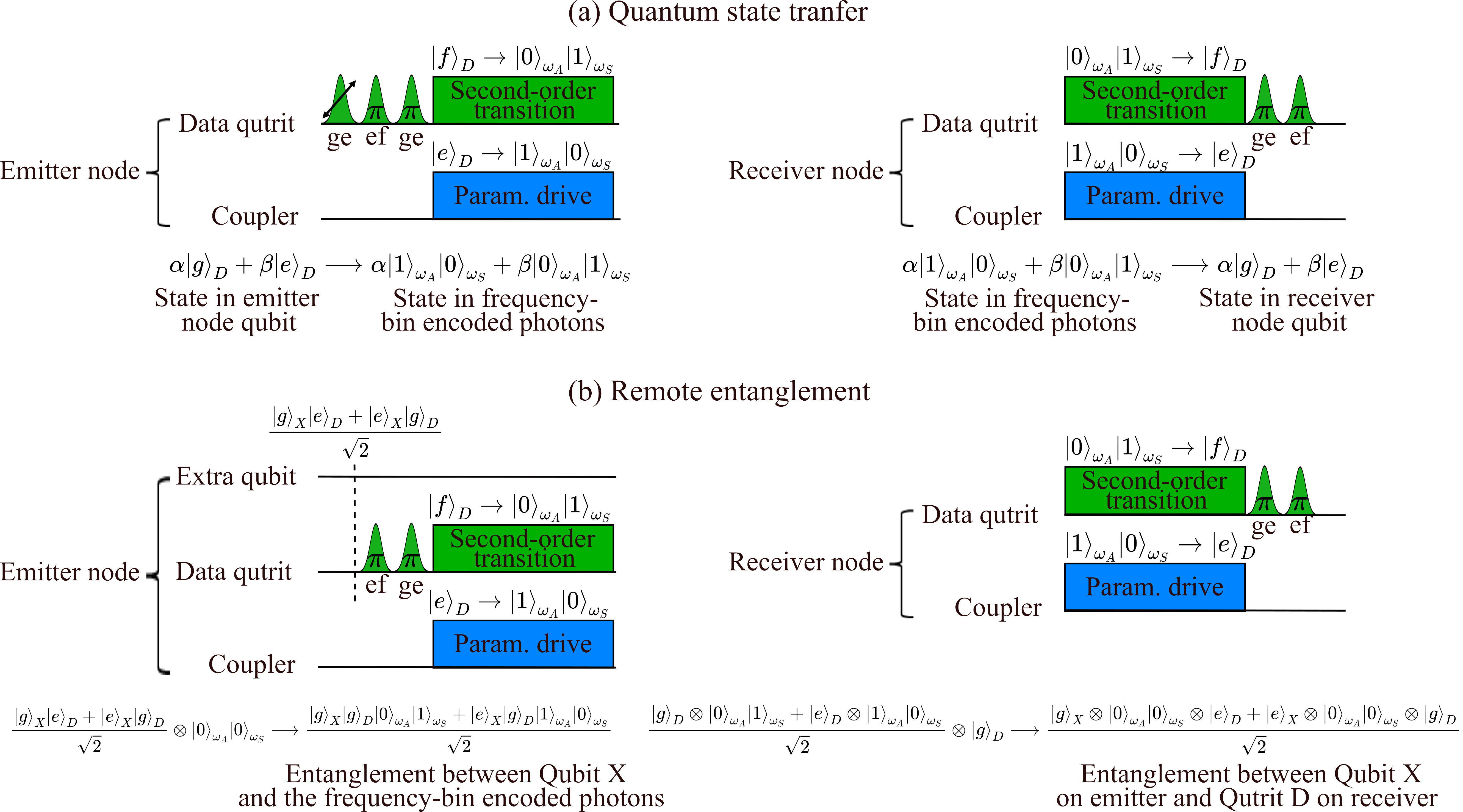}
\caption{The complete process of quantum state transfer and remote entanglement involves both the emitter and receiver nodes of the two quantum processors. In this work, we have experimentally implemented quantum state transfer at the emitter node. The remaining steps can be implemented similarly by simultaneously applying parametric and second-order-transition drives.}
\label{whole_process}
\end{figure*}

The frequency-bin-encoded photonic modes generated by our system have potential applications in quantum state transfer and establishing entanglement between two remote quantum processors, with the capability of detecting photon loss during transmission. While our experiment primarily demonstrates the generation of the frequency-bin photon, here we illustrate the complete process, which includes both the generation and the subsequent reabsorption of the photon by the receiving quantum processor.

The process of transferring quantum states between the emitter and receiver devices in our proposed frequency-bin-encoding-based distributed quantum computing system is shown in \figpanel{whole_process}{a}. As described in the main text, the emission of a frequency-bin-encoded photon begins by bringing the arbitrary superposition state $\alpha \left | g \right \rangle + \beta\left | e \right \rangle $ on Qutrit D into the state $\alpha \left | e \right \rangle + \beta\left | f \right \rangle $ using $\pi_{ef}$ and $\pi_{ge}$ pulses. Following this, the parametric drive and second-order-transition drive are applied simultaneously to emit the two frequency-bin-encoded photonic modes. At the receiver node, implemented with an identical quantum chip, the frequency-bin-encoded photonic modes are reabsorbed through the same process as the emission, but in reverse order. If any of the two frequency-bin photonic modes is lost before being absorbed,  the state of Qutrit D at the receiver will end up in the second excited state $\left | f  \right \rangle $ due to the last two $\pi$-pulses at the receiver. This is described in Table~\RNum{1} in the main text. 

The distributed quantum computing system facilitates the transfer of information between quantum processors. Typically, when using this system, the emitter and receiver nodes are coupled to extra quantum processors for computing large quantum tasks. To achieve entanglement between the qubits on the emitter and receiver devices, we can utilize the qubit in the quantum processor as an auxiliary qubit [\figpanel{whole_process}{b}]. At the emitter node, an auxiliary qubit (denoted as X) from the quantum processor is coupled to the distributed quantum computing system. The entanglement between Qubit X and Qutrit D is transferred to that between Qutrit D and the emitted frequency-bin-encoded photonic modes. At the receiver node, this establishes entanglement between Qubit X at the emitter node and Qutrit D at the receiver node. Finally, if needed, a parametric drive is used to swap the state into another Qubit X in the quantum processor at the receiver node, thereby entangling Qubit X across both nodes. Similar to quantum state transfer, if the photon is lost before re-absorption, we can detect it since the qubit at the receiver will end up in the second excited state $\left | f  \right \rangle $.

\nocite{*}

\bibliography{supplementary}